# Modeling of Human Body-coupled Electric Field Interference in Unshielded Ultra-Low Field MRI

**Authors**

Jiali He,[1] Yamei Dai,[1] Sheng Shen,[2,3] Jiamin Wu,[4] Zheng Xu[1*]

**Affiliations**

1. School of Electrical Engineering, Chongqing University, Chongqing 400044, China
2. A.A. Martinos Center for Biomedical Imaging, Massachusetts General Hospital, Boston, MA, USA.
3. Harvard Medical School, Boston, MA, USA.
4. Shenzhen Academy of Aerospace Technology, Shenzhen 518057, China.

**\*Correspondence to:**

Zheng Xu, Ph.D., Prof.
The school of Electrical Engineering
Chongqing University, Chongqing, China.
Email: xuzheng@cqu.edu.cn

**Abstract:** Portable ultra-low field MRI (ULF-MRI) systems operated in unshielded environments are susceptible to electromagnetic interference (EMI). Subject presence in the imaging region will lead to substantial noise increases, yet the dominant coupling mechanism remains insufficiently characterized. We develop a lumped-parameter circuit model of the coupled environment-body-receiver system. The model indicates that ambient time-varying electric fields induce a body common-mode potential, which is converted into differential-mode noise through capacitive imbalance between the head and the receive-coil terminals, yielding strong dependence on subject position and geometry. Circuit analysis, simulations, and controlled experiments support the model, with predicted imbalance consistent with measured noise variations. Guided by this mechanism, we implement a capacitive low-impedance bypass to clamp the body potential, achieving an approximately 3.5-fold SNR improvement on a 50 mT prototype. The proposed model offers a compact circuit-based tool for analyzing and mitigating human body-coupled electric-field interference in portable ULF-MRI.

## INTRODUCTION

Ultra-low-field magnetic resonance imaging (ULF-MRI) has attracted increasing attention in recent years due to its low cost, the absence of electromagnetic shielding requirements, and favorable portability [1]–[11]. These characteristics make ULF-MRI particularly promising for bedside diagnosis, emergency medicine, and healthcare delivery in resource-limited or remote settings [12]. However, ULF-MRI systems typically operate under low static magnetic fields (generally below 100 mT) and in non-shielded environments, where they inherently suffer from both weak intrinsic signal strength and pronounced susceptibility to external electromagnetic interference (EMI). The combined effect of these factors constitutes a primary limitation on the achievable signal-to-noise ratio (SNR) in ULF-MRI [7].

In non-shielded ULF-MRI systems, external EMI couples into the system mainly through conducted, magnetic-field, and electric-field pathways, as shown in Fig. 1 [13]. Conducted interference enters via power or ground lines and can be mitigated by filtering and grounding optimization [14]–[16]. Magnetic-field coupling results from electromagnetic induction, whereby low-frequency environmental fields induce differential-mode voltages in the receive coil. In addition, ferromagnetic yokes and structural components can change the internal interference magnetic-field distribution, leading to anisotropic attenuation and, in multichannel systems, channel-dependent SNR imbalance, as analyzed in our previous work [17].

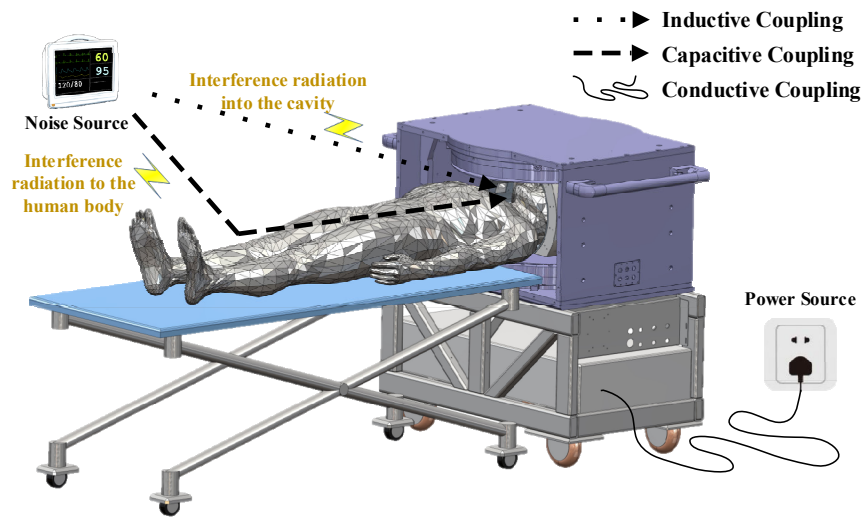

**Fig. 1.** Schematic illustration of EMI coupling pathways in an ultra-low-field MRI system.

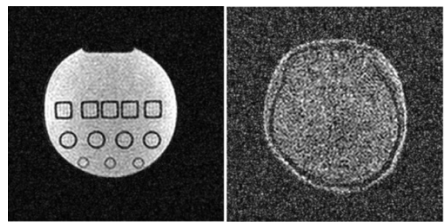

**Fig. 2.** Comparison between water phantom and brain imaging results acquired under identical experimental conditions.

Electric-field coupling is particularly pronounced in non-shielded ULF-MRI systems, as shown in Fig 2. Numerous studies have qualitatively reported a substantial noise increase when a subject enters the imaging volume [18]–[20]. The human body is directly exposed to the ambient electromagnetic environment and effectively acts as a coupling pathway between external EMI and internal system components, allowing external electric-field noise to couple into the centrally located RF receive coil.

From an engineering perspective, existing solutions can be grouped into two broad categories. The first relies on passive shielding to isolate the system from environmental EMI, often at the expense of portability and, in some cases, with increased discomfort for claustrophobic patients [21]–[23]. The second comprises active noise cancellation and signal/image-domain processing methods, such as noise-channel modeling [13], [24]–[27] and AI-based approaches [3], [9], [28]–[32] (e.g., deep learning). Although these methods can be effective, their performance often hinges on the accuracy of the assumed noise model and/or the availability of representative training data.

More generally, EMI mitigation in ULF-MRI remains largely "result-oriented," prioritizing post hoc suppression after interference has entered the receive chain rather than quantitatively characterizing the physical coupling pathways by which environmental noise reaches the RF receive coil. In particular, the human body in imaging is typically described only qualitatively as an "antenna effect," and a quantitative environment–body–system coupling model that explicitly treats the body as a dominant pathway is still lacking. As a result, in bedside settings, the mechanisms, coupling strength, and dominance conditions under which environmental RF noise couples into the receive coil remain insufficiently understood. Establishing such a model is therefore essential for developing physically interpretable, real-time mitigation strategies that avoid bulky shielding.

Motivated by these considerations, this work focuses on electric-field coupling and proposes an environment–human body–receiver modeling framework for ULF-MRI, complementing previously reported magnetic-field coupling models [17]. A multi-node lumped-element equivalent circuit model tailored for ULF-MRI is introduced to describe the transmission pathways of electric-field-related EMI and to analyze key sensitive parameters governing coupling strength. Based on this model, intervention and suppression strategies targeting capacitive coupling pathways are further discussed, with the aim of reducing electric-field-induced noise in the receive chain. The proposed framework provides a reusable modeling approach and theoretical basis for analyzing and mitigating EMI in portable ULF-MRI systems.

## METHODS

### A. Quasi-Static Capacitive Coupling

In unshielded bedside environments, the surrounding power infrastructure, lighting, and electronic devices act as distributed sources of time-varying electric fields. When a subject is placed in the imaging region, the body can be regarded as a conductive object covered by a thin insulating layer, so external electric fields induce redistribution of surface charges and a fluctuating body potential. This potential drives displacement currents through unavoidable parasitic capacitances among the body, the RF receive coil, and nearby metallic structures, providing a primary pathway for electric-field interference to enter the receive chain [33]–[36]. At the ULF-MRI operating frequency(2.23MHz) considered here, the interaction is dominated by quasi-static electric-field coupling; therefore, the environment–body–coil interaction can be modeled using an equivalent capacitive network in the subsequent noise-channel formulation [37].

### B. *Equivalent-Circuit Modeling of the Brain–Receive-Coil Interaction*

The preceding analysis shows that, in typical bedside scenarios, the electromagnetic environment around the human body is dominated by the reactive near field, where low-frequency electric-field coupling is governed by quasi-static charge distributions. Under these conditions, body–coil interactions can be naturally described using a capacitive coupling framework [33]–[37]. Accordingly, this section outlines the modeling assumptions for brain–coil electric-field coupling and develops a lumped-element capacitive circuit model to analyze the associated noise mechanism.

In the representative head ULF-MRI configuration (Fig. 3), saddle or solenoidal receive coils surround the head; taking the solenoidal coil as an example, anatomical constraints lead to partial axial loading of the coil, with remaining sections exposed to air or low-permittivity media. This partial-filling effect produces pronounced axial nonuniformity in the electromagnetic environment, which constitutes a key physical feature of human-mediated electric-field coupling.

To capture this spatially nonuniform coupling behavior, Fig. 3(b) presents a distributed-parameter model of the solenoidal coil and brain based on the partial element equivalent circuit (PEEC) method [38]. In this representation, the coil consists of *n* turns, each subdivided into *m* discrete elements connected in series. The *j*-th element of the *i*-th turn is characterized by its intrinsic distributed impedance, including resistance $R_{ij}$ and inductance $L_{ij}$, as well as a coupling capacitance $C_{ij}$ between the coil element and the brain. Owing to spatially varying geometric separation between different coil segments and the head, these coupling capacitances are not uniform but vary systematically along the coil axis. For the discrete node indexed by *i\*m+j*, the corresponding coupling capacitance $C_{ij}$ can be defined by a line integral along the conductor path.

$$C_{ij}(x_n) = \oint_{\Delta l_n} \frac{\epsilon(s) \cdot w}{d(s)} ds \tag{1}$$

where *w* denotes the effective conductor width, *d(s)* represents the local separation between the conductor and the brain surface, and *ε* is the effective relative permittivity along the integration path.

Based on the aforementioned physical partitioning, the discretized coil nodes can be classified into a proximal loaded region $\mathcal{N}_{prox}$ and a distal unloaded region $\mathcal{N}_{disk}$. Due to substantial differences in both dielectric properties and geometric conditions, the corresponding coupling capacitances in these two regions satisfy the inequality given

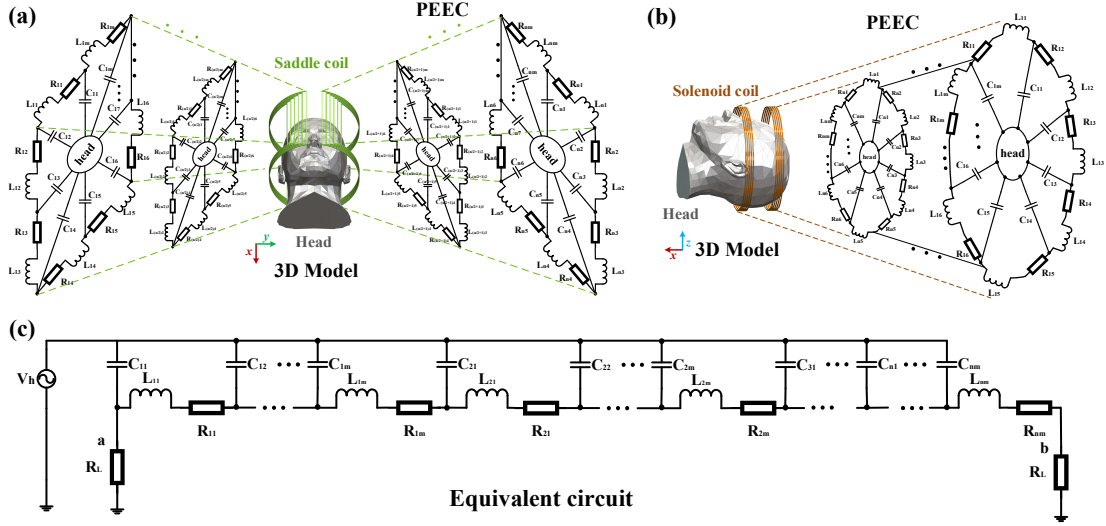

**Fig. 3.** Distributed-parameter models of the brain–coil system. (a) Saddle coil with the brain. (b) Solenoidal coil with the brain. (c) Corresponding equivalent circuit model.

in (2).

$$\forall k \in \mathcal{N}_{prox}, \forall l \in \mathcal{N}_{dist} \Rightarrow C_k > C_l \qquad (2)$$

This inequality captures the fundamental asymmetry of the brain–coil coupling and forms the physical basis for the subsequent development of the equivalent model.

While the distributed formulation captures local brain–coil coupling, this study focuses on the equivalent noise behavior at the coil terminals induced by the body-related common-mode potential rather than the internal voltage distribution. Accordingly, the distributed coupling network is reduced to a two-port lumped capacitive model. The Appendix rigorously derives this equivalence, showing that the effective terminal capacitances depend on both coil geometry and the spatially weighted distribution of coupling capacitances.

### C. *System-Level Equivalent-Circuit Model of the Environment–Body–Receive Chain*

Based on the preceding analysis, differential-mode interference arises primarily from nonuniform brain loading along the receive coil. Variations in the filling factor—axial for solenoidal coils and transverse for saddle coils—produce unequal effective coupling capacitances at the coil terminals, enabling conversion of common-mode noise into differential-mode interference. To capture the resulting interactions among the environment, the human body, and the receive coil, a unified system-level equivalent circuit model is therefore required.

Fig. 4(b) shows an integrated capacitive coupling model of the environment–body–coil system abstracted from the bedside scenario in Fig. 4(a). Environmental EMI is modeled as an equivalent source $V_E$ coupled to the body through $C_{EH}$. $V_E$ lumps the net contribution of ambient, spatially distributed low-frequency ***E***-field sources (e.g., power lines and lighting systems) into an effective excitation. Because the body is not an ideal ground, the finite body-to-ground impedance—set by $C_{Hg}$ and the effective

body–bed–ground capacitances $C_{Hn}$ and $C_{ng}$—establishes a body common-mode potential $V_h$. Nodes *a* and *b* represent the two terminals of the coil, which consists of an inductance $L_{coil}$ and a resistance $R_{coil}$ in series. The coil is connected to the downstream preamplifier (with input impedance $Z_L$) via a π-type matching network and a coaxial cable. The effective brain-to-terminal coupling capacitances $C_{Ha}$ and $C_{Hb}$ form the main paths that convert $V_h$ into differential-mode interference at the coil ports, while $C_{ag}$ and $C_{bg}$ account for terminal-to-ground parasitic capacitances; their equivalent representations are detailed in the appendix.

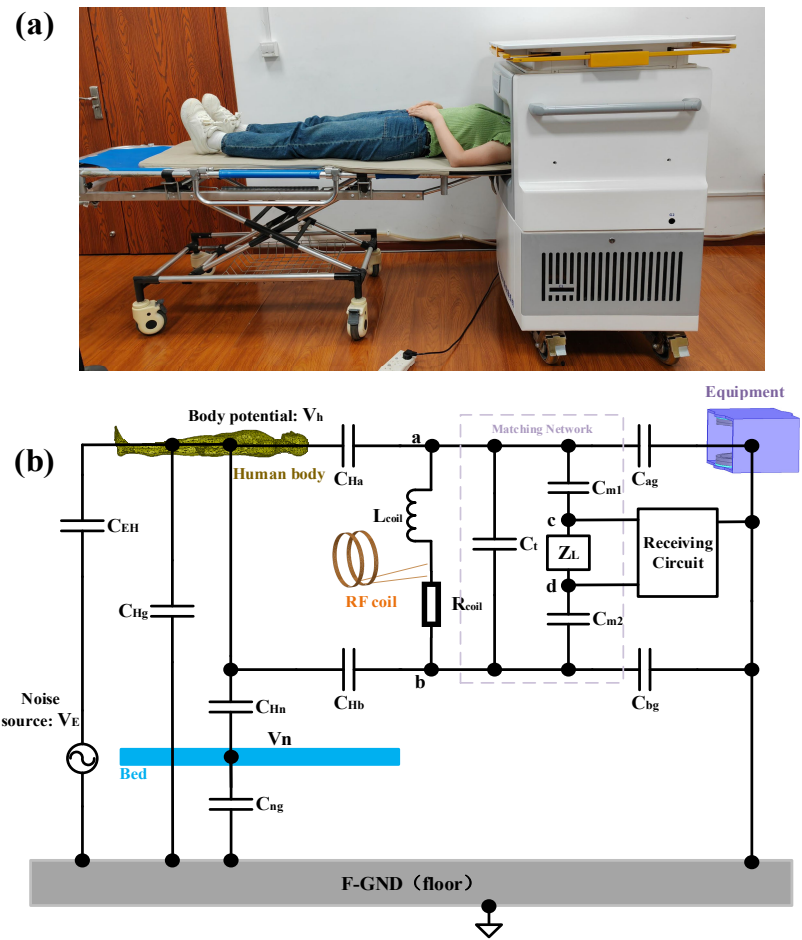

**Fig. 4.** (a) Imaging Scenario of the Non-Shielded ULF-MRI Platform. (b) Equivalent circuit model of human-mediated electric-field interference coupling.

The common-mode body potential $V_h$ acts as the effective driving source of the interference system, with its amplitude determined by the noise-source voltage $V_E$ through the associated voltage-divider network. Because the weak body-to-coil coupling capacitances (on the order of picofarads) exhibit very high impedances—significantly larger than the impedance of the body-to-ground return path—the loading effect of the receive coil can be neglected. Under this assumption, the body potential $V_h$ can be approximated as:

$$V_h \approx V_E \cdot \frac{C_{EH}}{C_{EH} + C_{HG}} \tag{3}$$

We first define the admittances of the individual circuit branches, starting with the admittance of the noise-injection path:

$$Y_{EH} = j\omega C_{EH}, \quad Y_{HG} = j\omega C_{HG} \tag{4}$$

Here, $C_{HG}$ denotes the total effective body-to-ground capacitance, incorporating both the direct body-to-ground capacitance $C_{Hg}$ and the capacitance associated with the bed-mediated grounding path.

$$C_{HG} = C_{Hg} + \frac{C_{Hn}C_{ng}}{C_{Hn} + C_{ng}} \tag{5}$$

The coupling admittances between the human body and coil terminals *a* and *b* are defined as $Y_{Ha}=j\omega C_{Ha}$ and $Y_{Hb}=j\omega C_{Hb}$, respectively. The parasitic admittances from the coil mechanical enclosure to ground at terminals a and b are denoted by $Y_{ag}=j\omega C_{ag}$ and $Y_{bg}=j\omega C_{bg}$. The coil resonant circuit, consisting of $L_{coil}$, $R_{coil}$, and the tuning capacitor $C_t$, together with the matching network branch $C_m$, is connected in parallel across terminals *a* and *b*. The resulting total shunt mutual admittance is defined as $Y_X$, given by:

$$Y_X = \frac{1}{R_{coil} + j\omega L_{coil}} + j\omega C_t + \frac{j\omega C_m}{2 + j\omega C_m Z_L} \tag{6}$$

The third term represents the admittance of the series branch formed by the matching capacitor $C_m$ and the load $Z_L$, arising from the floating condition of the output nodes *c* and *d*.

According to Kirchhoff's current law (KCL), the following relations can be written for nodes *a* and *b*:

$$\begin{cases}(V_a - V_h)Y_{Ha} + (V_a - V_b)Y_X + V_a Y_{ag} = 0 \\ (V_b - V_h)Y_{Hb} + (V_b - V_a)Y_X + V_b Y_{bg} = 0\end{cases} \tag{7}$$

Rearranging the above equations yields the standard linear matrix form **AV=I**:

$$\begin{bmatrix} Y_{aa} & -Y_X \\ -Y_X & Y_{bb} \end{bmatrix} \begin{bmatrix} V_a \\ V_b \end{bmatrix} = V_h \begin{bmatrix} Y_{Ha} \\ Y_{Hb} \end{bmatrix} \tag{8}$$

Here, the diagonal elements $Y_{aa}=Y_X+Y_{ag}+Y_{Ha}$ and $Y_{bb}=Y_X+Y_{bg}+Y_{Hb}$ represent the self-admittances of the corresponding nodes. The analytical solution for *Vab* is obtained as:

$$V_{ab} = V_h \frac{Y_{Ha}Y_{bg} - Y_{Hb}Y_{ag}}{D} \tag{9}$$

where:

$$D = Y_X(Y_{ag} + Y_{bg} + Y_{Ha} + Y_{Hb}) + (Y_{ag} + Y_{Ha})(Y_{bg} + Y_{Hb}) \tag{10}$$

To obtain the model with practical engineering relevance, two simplifying assumptions grounded in physical considerations are introduced. First, owing to the geometric symmetry of the RF coil and its enclosure, the parasitic admittances to ground at the two terminals can be assumed equal, i.e., $Y_{ag}=Y_{bg}$. Second, the coupling capacitances between the human body and the coil through air are much smaller than the capacitance between the metallic mechanical structure and ground, implying $Y_g \gg Y_{Ha}, Y_{Hb}$. Under these assumptions, (9) can be simplified to:

$$V_{ab} \approx V_h \cdot \frac{Y_{Ha}-Y_{Hb}}{2Y_X+Y_g} = V_h \cdot \frac{j\omega(C_{Ha}-C_{Hb})}{2[\dfrac{1}{R_{coil}+j\omega L_{coil}}+j\omega C_t+\dfrac{j\omega C_m}{1+j\omega C_m Z_L}]+j\omega C_g} \quad (11)$$

Finally, the input voltage $V_{cd}$ of the downstream circuitry is determined by the voltage-division effect of the matching network, corresponding to the detected noise signal:

$$V_{cd} = V_h \cdot Z_{total} \cdot j\omega(C_{Ha}-C_{Hb}) \quad (12)$$

Where,

$$Z_{total} = \frac{1}{2[\dfrac{1}{R_{coil}+j\omega L_{coil}}+j\omega C_t+\dfrac{j\omega C_m}{1+j\omega C_m Z_L}]+j\omega C_g} \cdot \left(\frac{j\omega C_m Z_L}{2+j\omega C_m Z_L}\right) \quad (13)$$

Equation (12) indicates that human-mediated electric-field interference coupling is primarily governed by three factors: (i) Source—the common-mode body potential $V_h$, which depends on the strength of external interference and the associated voltage-divider network; (ii) Conversion mechanism—the imbalance in coupling capacitances between the brain and the two coil groups, quantified as $\Delta C=C_{Ha}-C_{Hb}$; and (iii) Gain factor—the total impedance of the receive chain, including the matching network and the downstream circuitry.

### D. *Sensitivity Analysis of Noise Voltage*

Based on the equivalent circuit, (13) shows that the differential-mode noise voltage $V_{cd}$ is governed by three system parameters. Effective suppression without compromising imaging performance requires assessing which parameters are practically controllable under physical constraints.

First, $\Delta C$ quantifies the imbalance of the two coupling capacitances and thus the geometric asymmetry of the noise-transmission path. Although $\Delta C=0$ would ideally cancel differential-mode noise, $\Delta C$ is strongly perturbed by subject-specific anatomy and inevitable positioning variability, making picofarad-level balancing impractical and difficult to guarantee through coil geometry alone. Second, $Z_{total}$ sets the transfer gain from coupled noise to terminal voltage. While lowering $Z_{total}$ reduces the noise amplitude, electromagnetic reciprocity links receive sensitivity to coil Q and loop impedance, so decreasing $Z_{total}$ also reduces MR signal amplitude and yields little net SNR benefit.

With geometric balance and loop impedance both constrained, suppressing the body common-mode potential $V_h$ becomes the primary remaining degree of freedom. From the divider relation in (3), $V_h$ is determined by the ratio of the body-to-ground impedance to the environment-to-body coupling impedance. Accordingly, we introduce a capacitive grounding strategy (Fig. 5) that reshapes the body-to-ground network by placing an insulated conductive layer near the subject, effectively adding a large shunt capacitance $C_{HG}$ to ground in parallel. Because $C_{HG} \gg C_{EH}$, this provides a low-impedance bypass for common-mode currents and clamps the otherwise floating $V_h$ toward ground, suppressing differential-mode noise at its physical origin. Importantly, this approach is largely independent of the RF receive loop, enabling mitigation of electric-field-coupled interference without degrading geometric balance or reducing loop impedance.

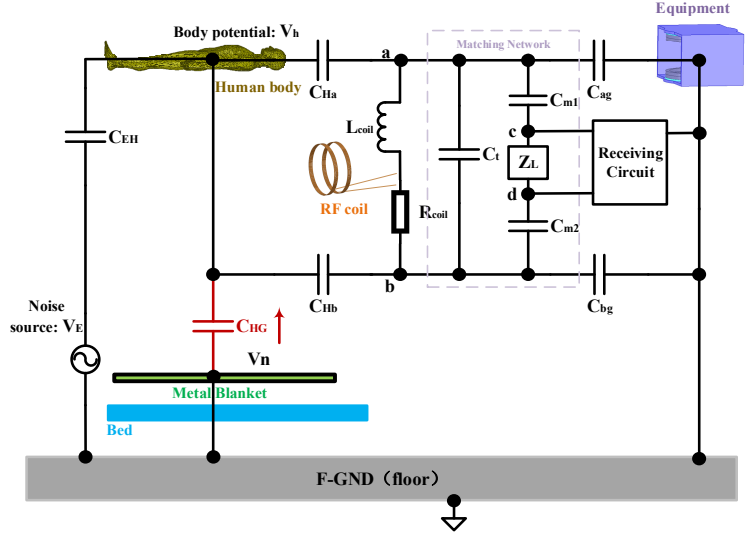

**Fig. 5.** Equivalent circuit model with an added grounded metallic blanket.

## EXPERIMENTS AND RESULTS

To validate the model, three types of simulations and experiments were performed. First, capacitive imbalance between the head and two receive-coil types was investigated by varying the head–coil relative position and analyzing the resulting noise changes. Second, the effect of the body-related potential $V_h$ was evaluated by fixing the head–coil position (thus keeping the coupling-capacitance difference constant) while varying the exposed area to modulate $V_h$; aluminum plates of different lengths were used to emulate this condition. Third, the effectiveness of a grounded metallic blanket was assessed by comparing the resulting noise levels and image SNR.

### A. *Experimental Setup*

The experimental platform used in this study was a self-developed, portable, non-shielded 50 mT ULF-MRI scanner, as shown in Fig. 6. To achieve portability and electromagnetic isolation, all electronic subsystems, including the gradient power amplifiers, radio-frequency (RF) power amplifier, spectrometer, and RF switching unit, were enclosed within an aluminum housing and integrated into the magnet frame, with the exception of the preamplifier. Signal acquisition was performed using an MR solution EVO spectrometer (MR Solutions, UK) with 8 receive channels and 1 transmit channel. Both the solenoidal and saddle receive coils were tuned and impedance-matched to 2.23 MHz.

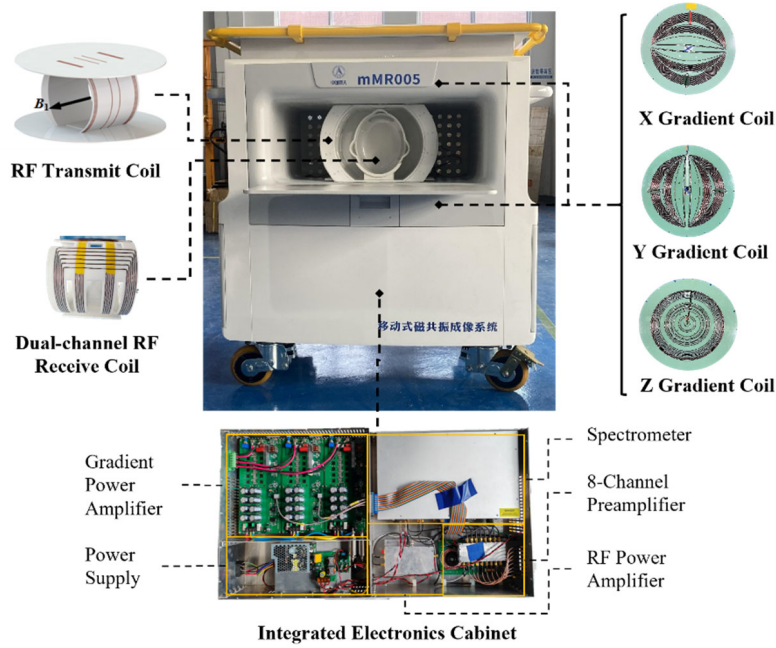

**Fig. 6.** Portable 50 mT ULF-MRI system.

The scanner was deployed inside a building located in an industrial park. During data acquisition, no restrictions were imposed on the external electromagnetic environment. Major sources of interference included nearby MRI systems, LED lighting infrastructure, and mobile phones. Unless otherwise specified, all experiments reported in this work were conducted under these conditions. All imaging experiments employed a three-dimensional gradient-echo (GRE) sequence, with an acquisition time of approximately 2 minutes per scan.

### B. *Noise Induced by ΔC*

This section validates the influence of the coupling-capacitance difference $\Delta C$ on common-mode noise conversion predicted by (12). Given the anatomical complexity of the human head and large inter-subject variability, we do not attempt to extract subject-specific absolute capacitance values; instead, we focus on how capacitive imbalance evolves with head–coil misalignment and how it governs the terminal noise amplitude.

#### *1) Solenoidal coil*

To enable trend-oriented analysis with physically grounded simplifications, the head was represented by a hemisphere–cylinder composite (Fig. 7(b)) that preserves the main volume while introducing axial asymmetry relevant to $\Delta C$. For the solenoidal receive coil, the discrete windings were approximated by two axially separated continuous conductive rings with identical effective coverage. This approximation is justified by electrostatic homogenization: in the quasi-static regime, fringe fields bridge inter-turn gaps, rendering tightly wound solenoids capacitively equivalent to a continuous conductor. Ansys Q3D simulations confirm this equivalence, showing less than a 5% difference in coil-to-ground coupling capacitance between the two models, thereby supporting the ring representation as an efficient surrogate for coil–head capacitive coupling.

Using these simplified models, coil–head coupling capacitances were computed (via a coaxial cylindrical-capacitor approximation) while incrementally translating the head model along the X-axis to emulate varying relative positions. Fig. 7(a) reports the coupling capacitances to the posterior and anterior coil segments and their difference $\Delta C$ (origin at the coil center, abscissa defined by the displacement of the head apex). A finite $\Delta C$ exists even at the nominally centered position due to intrinsic head asymmetry. As the head moves in the +X direction, the posterior coupling decreases first, whereas the anterior coupling remains relatively high, producing a pronounced rise in $\Delta C$; once the head moves beyond the effective coverage of the anterior segment, the anterior coupling also decreases and $\Delta C$ correspondingly drops, yielding a characteristic increase–decrease profile. Fig. 7(c) shows the measured terminal noise amplitude between nodes $a$–$b$, whose peak variation closely tracks $\Delta C$, thereby confirming that capacitive imbalance is the dominant driver for converting body-related common-mode excitation into differential-mode interference in this configuration.

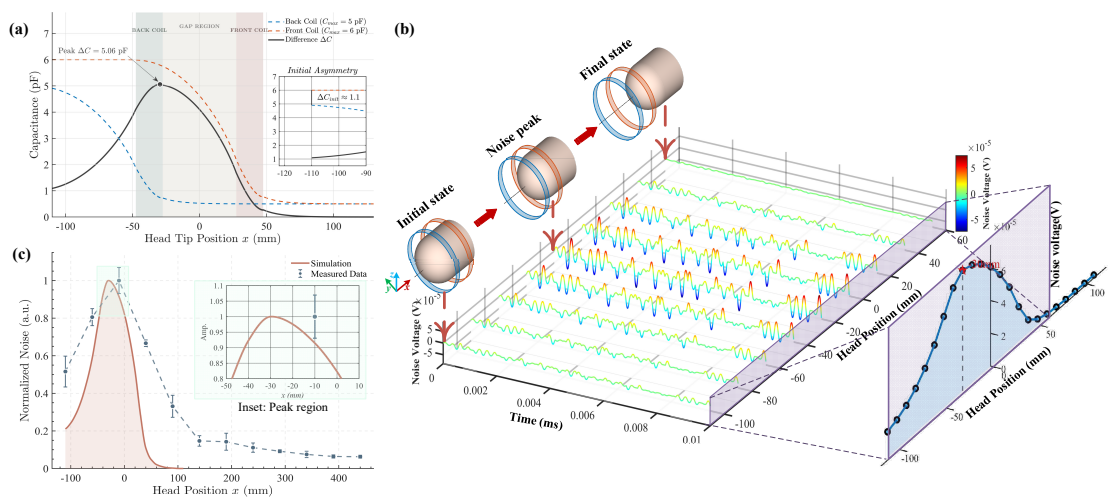

**Fig. 7.** Simulation and experimental validation of noise mechanisms induced by capacitive coupling imbalance. (a) Simulated coil–brain coupling capacitances ($C_{Ha}$, $C_{Hb}$) and their difference ($\Delta C$) as functions of head displacement along the positive X-axis; (b) simplified geometric model used in simulations, illustrating the relative configuration of the solenoidal coil (conductive ring bands) and the head model (hemisphere–cylinder composite). The extracted noise envelope and peak amplitude reflect the dynamic noise behavior as the head gradually moves out of the coil; (c) experimentally measured normalized noise amplitude in the solenoidal coil as a function of head position, demonstrating agreement between simulation predictions and measurements.

To corroborate the simulations, experimental measurements were performed with the constructed setup while keeping all other system components unchanged. The subject's head was gradually moved out of the solenoidal coil, and Fig. 7(c) shows that the measured normalized noise amplitude agrees well with the predicted trend. The experimental baseline is slightly higher than the simulated level, likely due to additional contributions (e.g., circuit thermal noise and environmental magnetic interference) that are not included in the simplified capacitive coupling model.

Representative imaging results under different head positions are shown in Fig. 8. When the head is centered in the solenoidal coil, the mean SNR within the white boxed region is ~5.5; with an offset head position, the mean SNR drops to ~2.3, corresponding to a ~2.4× reduction. Consistent with this quantitative metric, the reconstructed images and one-dimensional intensity profiles at matched anatomical locations exhibit a pronounced noise increase under the offset condition, with partial masking of the MR signal and degraded delineation of tissue boundaries.

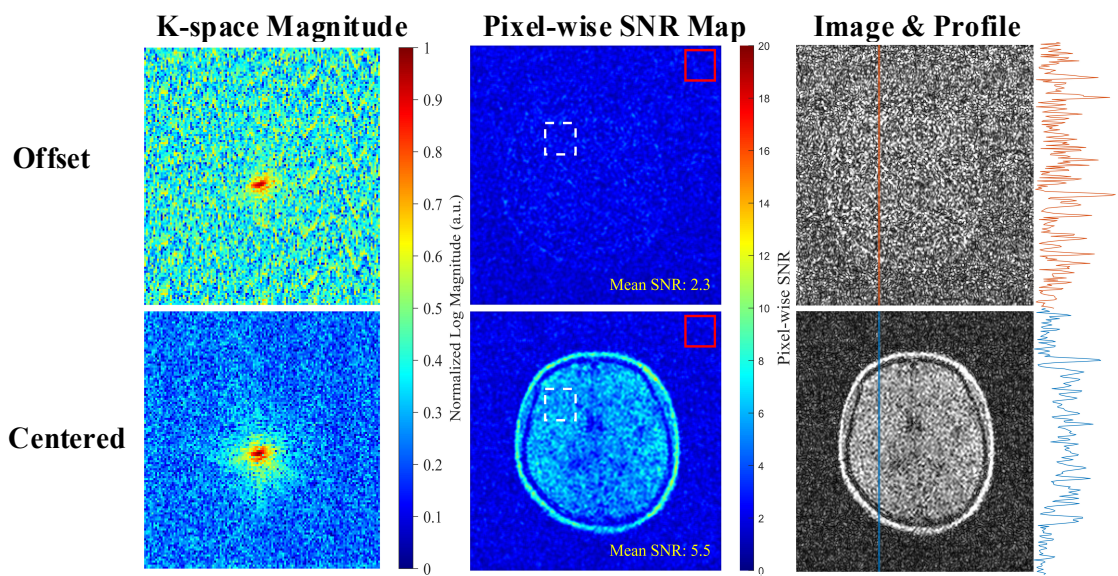

**Fig. 8.** Imaging comparison with the head centered and offset relative to the solenoidal coil. From left to right: logarithmic k-space magnitude spectra, pixel-wise SNR maps, and reconstructed images with corresponding one-dimensional intensity profiles.

*2)* Saddle coil

After establishing the capacitive-imbalance noise mechanism for the solenoidal coil, we experimentally validate the same mechanism for a saddle coil. Although the saddle geometry differs, differential-mode interference still originates from coupling-capacitance imbalance under a body-related common-mode potential; the main difference is that saddle coils are more sensitive to transverse (Y-axis) asymmetry. Therefore, to avoid redundancy, we omit detailed simulations and verify the predicted noise behavior directly via measurements.

Fig. 9 summarizes the measured noise trends under controlled asymmetry. In Fig. 8(a), varying the head azimuth shows minimum noise when the head is centered, whereas rotation toward either side progressively breaks symmetry and increases noise. Because head motion is limited within the saddle coil, we further used the subject's hands as surrogate loads (Fig. 8(b)). With single-hand loading, noise is minimized at the coil center and increases as the hand moves toward the edge; with symmetric two-hand loading, the noise is markedly reduced. Together, these results demonstrate that restoring symmetric loading compensates capacitance imbalance and confirms that the coupling-capacitance difference is the dominant driver of noise generation in the saddle-coil configuration.

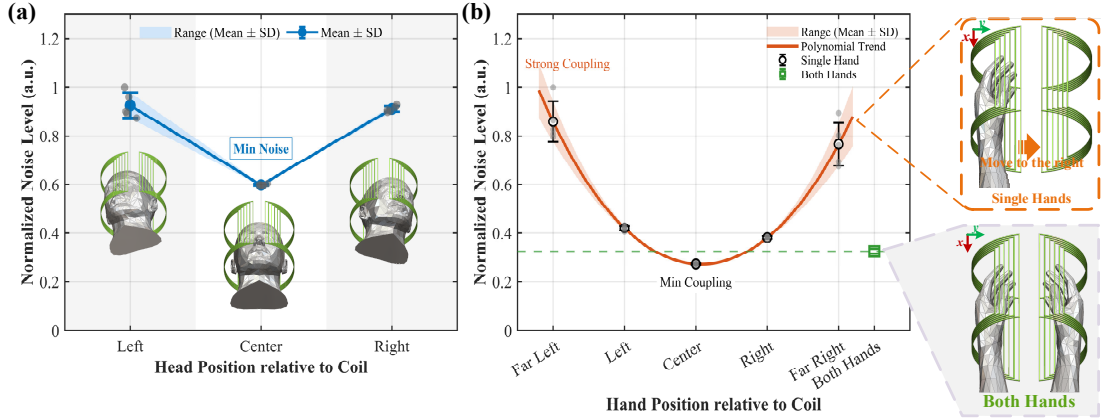

**Fig. 9.** Measured noise variation detected by the saddle coil as a function of object position. (a) Noise amplitude versus head azimuthal deviation (left–right rotation) relative to the saddle coil; (b) symmetry validation using a hand-based surrogate model, comparing noise levels under single-hand displacement and symmetric two-hand placement.

### C. *Noise Induced by $V_h$*

This section isolates the effect of the body common-mode potential $V_h$ on terminal noise, as predicted by (12), while keeping the coupling-capacitance imbalance $\Delta C$ unchanged. Because $V_h$ increases with stronger coupling to ambient electric fields, we modulate the effective common-mode excitation by varying the exposed length/area of a conductive object in free space.

To prevent subject motion and inter-subject geometry from altering head–coil coupling and $\Delta C$, we replace the subject with an aluminum-plate phantom that behaves as a stable floating conductor and allows repeatable positioning. The plate overlap inside the coil is fixed to maintain constant coil–load coupling (and $\Delta C$), and only the length extending outside the coil is varied from 50 to 200 cm to change environment coupling and $V_h$. Under this design, the measured noise variation is attributed primarily to changes in $V_h$ [18].

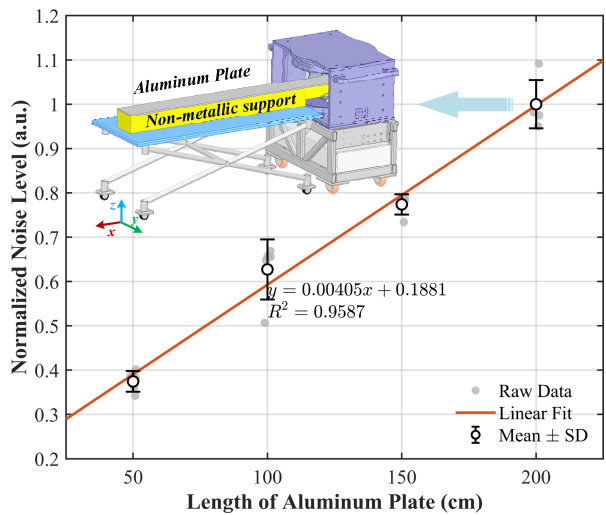

**Fig. 10.** Linear increase of terminal noise with aluminum-plate exposure length.

The exposed portion of a floating conductor in a non-shielded environment behaves like a monopole receiver whose ability to intercept the background electric field increases with length, which the proposed circuit model captures primarily as an increase in the common-mode potential $V_h$. As shown in Fig. 10, the noise amplitude measured at the coil terminals increases approximately linearly with the exposure length, consistent with the predicted linear dependence on $V_h$. These results support the interpretation that, once $\Delta C$ is fixed, variations in environment coupling strength dominate the modulation of system noise, validating the common-mode source term in the proposed equivalent model.

### D. Suppression Strategy

Fig. 11 summarizes the impact of introducing a capacitive grounding blanket on a non-shielded 50 mT ULF-MRI platform. In the frequency domain (Fig. 11(a), left and middle), the baseline logarithmic k-space magnitude spectra exhibit strong broadband, unstructured background components, indicating substantial common-mode interference from the surrounding environment. With the grounding blanket applied, this k-space clutter is markedly suppressed, and signal energy becomes more concentrated, demonstrating effective attenuation of interference components.

These improvements translate directly to image-level performance. Pixel-wise SNR maps show a pronounced SNR increase across the brain; within the white-boxed region, the mean SNR rises from 5.5 to 19.5 (≈3.5×). Consistent with this quantitative gain, reconstructed images and matched one-dimensional intensity profiles (Fig. 10(a), right) reveal a clear reduction in background fluctuations. Tissue contrast and boundary delineation are improved, while no noticeable loss of fine structural detail is observed, indicating that the suppression reduces electric-field-coupled interference without sacrificing image resolution.

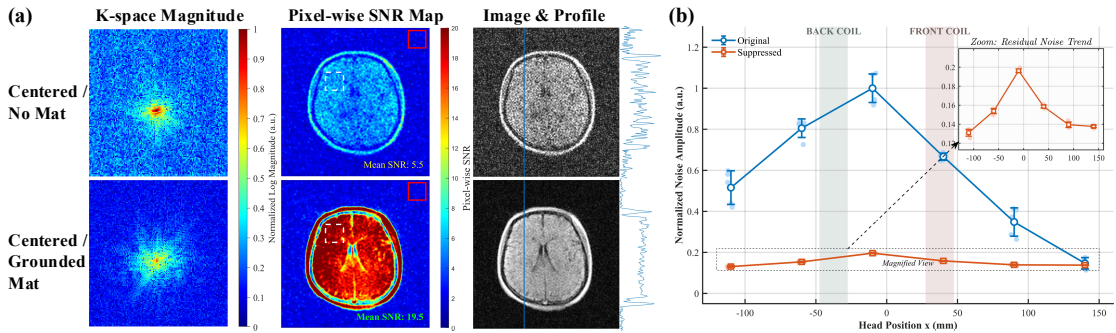

Fig. 11. Validation of the proposed suppression strategy. (a) Baseline and suppressed imaging: k-space spectra, SNR maps, and reconstructed images with intensity profiles, showing reduced background noise and improved SNR.(b) Noise amplitude versus axial head position without and with suppression, demonstrating effective attenuation of position-dependent interference.

To evaluate robustness to subject-position variability, Fig. 11(b) reports the noise amplitude as a function of head position along the axial direction ($x \in [-100\text{mm}, 150\text{mm}]$). Without suppression, noise shows strong position dependence and peaks near the coil center ($x=-10$mm), consistent with the previously analyzed imbalance-driven trend. With the grounding blanket, noise is suppressed across the entire range, with an overall reduction exceeding 80%; although a weak residual position dependence remains, its absolute magnitude is greatly diminished, demonstrating improved stability under practical bedside conditions.

**DISCUSSION**

This study investigated the physical origin of human body-couple noise and corresponding suppression strategies in portable ultra-low-field MRI systems operating in non-shielded environments. Experiments confirm that the human body provides a significant coupling pathway that elevates system noise and degrades image quality via electric-field coupling. The discussion focuses on (i) the dominant physical factors governing noise increments, (ii) a comparison of front-end suppression strategies for electric-field-coupled noise, and (iii) coil-geometry-dependent noise sensitivity.

*A. Dominant Contribution of Coupling-Capacitance Imbalance to Noise*

In the head-displacement experiment (Fig. 7(c)), two effects vary concurrently: the brain–coil coupling-capacitance imbalance $\Delta C$ changes with relative position, while the exposed body area outside the coil increases and may modulate the common-mode potential $V_h$. An order-of-magnitude estimate helps separate their contributions. When the head is displaced by ~8 cm, the effective exposure area increases by only ~4.7%, yet the measured noise nearly doubles (~100%). If the noise were mainly driven by $V_h$ variation, the change would be expected to be comparable to the exposure-area change. The substantially larger observed increment therefore indicates that, under this configuration, noise growth is dominated by variations in $\Delta C$, whereas exposure-induced changes in $V_h$ play a secondary role. This supports the conclusion that anatomical asymmetry and the resulting capacitive mismatch are major contributors to noise variation in practical non-shielded measurements.

*B. Categories of Electric-Field Noise Suppression*

Existing front-end strategies for mitigating human-mediated electric-field-coupling noise can be broadly divided into two categories: interruption of the electric-field coupling pathway and suppression of the body-related common-mode potential via bypass mechanisms, the latter being adopted in this work.

The first category aims to directly block coupling paths. Typical examples include Faraday cage structures, which effectively isolate the system from external EMI but substantially compromise portability and openness in portable MRI [21]–[23]. Other approaches insert electric-field shielding layers—such as comb-like or mesh structures—between the receive coil and the head to reduce brain–coil capacitive coupling ($C_{Ha}$/$C_{Hb}$) [20]. However, these methods generally increase coil–object separation, lowering the filling factor and unavoidably reducing signal sensitivity.

The second category suppresses electric-field-coupling noise through bypass pathways, such as direct grounding of the subject or [18], [19], [24], [39], as proposed here, introducing a large effective body-to-ground capacitance ($C_{HG}$) to reduce the common-mode body potential $V_h$. Compared with path-interruption approaches, capacitive bypass strategies better preserve system openness and portability while minimally affecting the filling factor. From an equivalent-circuit viewpoint, their effectiveness is governed by a voltage-divider relationship between the environment–body coupling capacitance ($C_{EH}$) and the body-to-ground capacitance ($C_{HG}$). Nevertheless, practical constraints on coupling area and dielectric properties limit the achievable magnitude of $C_{HG}$, such that the body common-mode potential $V_h$ can be attenuated but not fully eliminated. As shown in Fig. 11(b), residual noise remains after bypass suppression and still exhibits capacitive-coupling characteristics, including position dependence. Therefore, applications requiring high imaging sensitivity often combine such front-end strategies with backend signal processing or software-based denoising to further reduce residual interference [3], [13], [24].

### C. Coil-Geometry-Dependent Noise Sensitivity

The geometry of the receive coil has a pronounced impact on its sensitivity to both human-mediated coupling noise and environmental magnetic interference. Solenoidal coils inherently offer reduced sensitivity to magnetic-field interference [17]; however, as analyzed in this work, the pronounced anterior–posterior anatomical asymmetry of the human body readily introduces front–back capacitive mismatch when a subject is present, which can become a major pathway for human-mediated electric-field coupling noise.

In contrast, saddle coils extend laterally, and the human body generally exhibits higher left–right anatomical symmetry, leading to smaller capacitance imbalance and reduced sensitivity to human electric-field coupling noise. Nevertheless, as shown in Fig. 9 and in prior studies, saddle coils typically exhibit higher baseline noise in non-shielded environments. This behavior has been attributed to the dominance of transverse magnetic noise within the imaging enclosure and the higher coupling efficiency of saddle-coil geometries to such fields [17]. These differences indicate that, in portable ultra-low-field MRI systems, receive-coil selection inherently involves a trade-off between competing electric- and magnetic-noise mechanisms.

The present study focuses on a single-channel receive configuration to isolate the body-mediated electric-field coupling mechanism and to enable controlled experimental validation. Extending the framework to multichannel arrays will require accounting for inter-channel coupling, which can introduce additional common-mode to differential-mode conversion paths. Future work will therefore explore (i) improving receive-structure symmetry and incorporating distributed or tunable capacitive compensation to reduce $\Delta C$-driven conversion at the source, and (ii) combining the passive bypass with active noise sensing or backend processing to further improve performance in unshielded portable ULF-MRI while preserving openness and portability.

### CONCLUSION

This study establishes an equivalent-circuit model for understanding and mitigating human body-coupled electric-field interference in unshielded ULF-MRI. First, we built a lumped-parameter capacitive model of the environment–body–coil system that explicitly captures the noise transmission path to the receive-coil terminals. Second, the model identifies the key conversion mechanism: ambient electric fields induce a body common-mode potential, which is transformed into differential-mode interference primarily through coupling-capacitance imbalance. Finally, leveraging this insight, we implement a capacitive bypass that provides a low-impedance return path to ground to clamp the body potential, achieving substantial noise suppression and a ~3.5× SNR improvement on a self-developed 50 mT prototype. These results provide a practical, physically interpretable basis for front-end EMI mitigation and EMC-oriented design in portable ULF-MRI systems.

**APPENDIX**

*A. Lumped-Parameter Equivalent Model*

This section details the mathematical derivation that reduces a multi-node partial-element equivalent circuit model to a two-capacitance lumped-parameter representation. As shown in Fig. 3(c), the receive coil is discretized into Q series-connected differential elements. The q-th element (q=1,…, Q) is characterized by a local series impedance zq and a partial capacitance Cq to the human body. The current injected into the coil at the q-th node, denoted as i(q), is expressed as:

$$i(q) = j\omega C_q (V_h - v_q) \tag{14}$$

Given that this study operates in a weak-coupling, high-impedance regime, the induced noise potential is much smaller than the source common-mode potential:

$$i(q) \approx j\omega C_q V_h \tag{15}$$

Assuming a uniform coil structure with impedance linearly distributed along the conductor length, the path impedances from the *q*-th node to ports *a* and *b* can be expressed, respectively, as:

$$Z_{q \to a} \approx \frac{q-1}{Q-1} Z_{coil}, \quad Z_{q \to b} \approx \frac{Q-q}{Q-1} Z_{coil} \tag{16}$$

Here, $Z_{coil}$ denotes the total impedance of the coil. Under the large-number approximation (Q≫1), Q−1≈Q. According to the current-division principle, the portion of the injected current *i(q)* that returns to port *a*, denoted as $i_a(q)$, is inversely proportional to the corresponding impedance:

$$i_a(q) = i(q) \cdot \frac{Z_{q \to b}}{Z_{q \to a} + Z_{q \to b}} = i(q) \cdot \frac{Q-q}{Q} \tag{17}$$

Similarly, the current component that returns to port B, denoted as ib(q), is given by:

$$i_b(q) = i(q) \cdot \frac{Z_{q \to a}}{Z_{q \to a} + Z_{q \to b}} = i(q) \cdot \frac{q}{Q} \tag{18}$$

Following the principle of *port current conservation*, the total displacement current observed at port *a* in the macroscopic lumped model, denoted as $I_{micro\_A}$, must be strictly

equal to the sum of the current contributions diverted from all nodes in the microscopic model:

$$I_{micro\_A} = \sum_{q=1}^{Q} i_A(q) = j\omega V_h \cdot \left[\frac{1}{Q}\sum_{q=1}^{Q} C_{qh}(Q-q)\right] \quad (19)$$

In the macroscopic equivalent circuit, the coupling to ground at port $a$ is modeled as a single lumped capacitance $C_{Ha}$, and the resulting displacement current is defined as:

$$I_{macro\_A} = j\omega C_{Ha} V_h \quad (20)$$

By combining the above two expressions and eliminating the common factors, a rigorous definition of the lumped capacitances $C_{Ha}$ is obtained as:

$$C_{Ha} = \frac{1}{Q}\sum_{q=1}^{Q} C_{qh}(Q-q), \quad C_{Hb} = \frac{1}{Q}\sum_{q=1}^{Q} C_{qh} \cdot q \quad (21)$$

This expression shows that the macroscopic capacitance is a position-weighted sum of the distributed capacitances $C_q$. To interpret (21), a two-cluster solenoidal-coil model is adopted: the first $Q/2$ elements adjacent to the forehead form a front cluster with a larger unit capacitance $C_{front\_unit}$, while the remaining elements constitute a rear cluster with a smaller unit capacitance $C_{rear\_unit}$. Substituting this model into (21) yields:

$$\begin{aligned} C_{Ha} &= \frac{1}{Q}\left[\sum_{q=1}^{Q/2} C_{front\_unit}(Q-q) + \sum_{q=Q/2+1}^{Q} C_{rear\_unit}(Q-q)\right] \\ &\approx \frac{3}{4}\left(\frac{Q}{2}C_{front\_unit}\right) + \frac{1}{4}\left(\frac{Q}{2}C_{rear\_unit}\right) \\ &\approx 0.75 C_{TF} + 0.25 C_{TR} \end{aligned} \quad (22)$$

Here, $C_{TF}$ and $C_{TR}$ denote the total physical capacitances of the front and rear clusters, respectively. By the same reasoning, one obtains:

$$C_{Hb} \approx 0.25 C_{TF} + 0.75 C_{TR} \quad (23)$$

The derivation shows that asymmetric loading can produce significant differences between the terminal capacitances $C_{Ha}$ and $C_{Hb}$ even for a geometrically symmetric coil. When the forehead is close to the coil ($C_{TF} > C_{TR}$), $C_{Ha}$ is dominated by the term $0.75\ C_{TF}$, explaining the origin of $C_{Ha} > C_{Hb}$. The same analysis applies to saddle coils.

### ACKNOWLEDGMENT

The author would like to thank Dr. Lei Yang, Dr. Yuxiang Zhang, and Dr. Cai Wan for their helpful discussions on the experiment.